\documentclass[apj]{emulateapj}
\usepackage{mathptmx}

\newcommand  \acc     {\ifmmode {\rm km\,s}^{-2} \else km\,s$^{-2}$\fi}
\newcommand  \kms      {\ifmmode {\rm km\,s}^{-1} \else km\,s$^{-1}$\fi}

\newcommand  \ergs     {\ifmmode {\rm ergs\,s}^{-1} \else ergs s$^{-1}$\fi}
\newcommand  \ergcms   {\ifmmode {\rm ergs\,cm}^{-2}\,{\rm s}^{-1}
                        \else ergs\,cm$^{-2}$\,s$^{-1}$\fi}
\newcommand  \ergcmsA  {\ifmmode{\rm ergs\,cm}^{-2}\,{\rm s}^{-1}\,{\rm\AA}^{-1}
                        \else ergs\,cm$^{-2}$\,s$^{-1}$\,\AA$^{-1}$\fi}
\newcommand  \ergcmsHz {\ifmmode{\rm ergs\,cm}^{-2}\,{\rm s}^{-1}\,{\rm Hz}^{-1}
                        \else ergs\,cm$^{-2}$\,s$^{-1}$\,Hz$^{-1}$\fi}
\newcommand  \phcms    {\ifmmode {\rm ph\,cm}^{-2}\,{\rm s}^{-1}
                        \else ph\,cm$^{-2}$\,s$^{-1}$\fi}
\newcommand  \phcmsA   {\ifmmode {\rm ph\,cm}^{-2}\,{\rm s}^{-1}\,{\rm\AA}^{-1}
                        \else ph\,cm$^{-2}$\,s$^{-1}$\,\AA$^{-1}$\fi}

\newcommand  \rblr     {$R_{\rm BLR}$}

\journalinfo{The Astrophysical Journal, 
629:1--11, 2005 August 10, astro-ph/0504484}
\slugcomment{Received 2004 December 13; accepted 2005 April 20}

\shorttitle{BLR-SIZE AND LUMINOSITY RELATION IN AGNS}

\shortauthors{KASPI ET Al.}

\begin{document}

\title{The Relationship Between Luminosity and Broad-Line Region Size in
Active Galactic Nuclei}

\author{
Shai~Kaspi,\altaffilmark{1,2} 
Dan~Maoz,\altaffilmark{1} 
Hagai~Netzer,\altaffilmark{1} 
Bradley M. Peterson,\altaffilmark{3} \\   
Marianne Vestergaard,\altaffilmark{4}     
and
Buell T. Jannuzi\altaffilmark{5}          
}

\altaffiltext{1}{School of Physics and Astronomy, Raymond and Beverly
Sackler Faculty of Exact Sciences, Tel-Aviv University, Tel-Aviv 69978,
Israel; shai@wise.tau.ac.il.}
\altaffiltext{2}{Physics Department, Technion, Haifa 32000, Israel.}
\altaffiltext{3}{Department of Astronomy, The Ohio State University, 140 West 18th Avenue, Columbus, OH 43210.}
\altaffiltext{4}{Steward Observatory, The University of Arizona, 933 North Cherry Avenue, Tucson, AZ 85721.}
\altaffiltext{5}{National Optical Astronomy Observatory, P.O. Box
26732, Tucson, AZ 85719.}

\begin{abstract}
We reinvestigate the relationship between the characteristic
broad-line region size (\rblr ) and the Balmer emission-line, X-ray,
UV, and optical continuum luminosities.  Our study makes use of the
best available determinations of \rblr\ for a large number of active
galactic nuclei (AGNs) from Peterson et al. Using their determinations
of \rblr\ for a large sample of AGNs and two different regression
methods, we investigate the robustness of our correlation results as
a function of data sub-sample and regression technique. Though small
systematic differences were found depending on the method of analysis,
our results are generally consistent. Assuming a power-law relation
\rblr\,$\propto L^{\alpha}$, we find the mean best-fitting $\alpha$ is
about $0.67\pm0.05$ for the optical continuum and the broad H$\beta$
luminosity, about $0.56\pm0.05$ for the UV continuum luminosity,
and about $0.70\pm0.14$ for the X-ray luminosity. We also
find an intrinsic scatter of $\sim 40$\% in these relations. The
disagreement of our results with the theoretical expected slope of
0.5 indicates that the simple assumption of all AGNs having on average
same ionization parameter, BLR density, column density, and ionizing
spectral energy distribution, is not valid and there is likely some
evolution of a few of these characteristics along the luminosity scale.
\end{abstract}


\section{Introduction}

There is mounting evidence that massive black holes reside in the
centers of most or all massive galaxy bulges. Understanding the
demographics and properties of these black holes will hopefully
clarify their roles in galaxy formation and evolution, the ionization
of the intergalactic medium, and more.  The masses of the central
black holes in nearby non-active galaxies have been measured using
stellar and gas kinematics  of the central regions (Tremaine et
al. 2002 and references therein). In active galactic nuclei (AGNs),
the technique of reverberation mapping (Blandford \& McKee 1982;
Peterson 1993; Netzer \& Peterson 1997) has been used to measure
the light-travel-time delay over which broad emission line flux
responds to continuum luminosity variations, and to thus deduce the
characteristic size of the broad-line region (BLR) around the central,
photoionizing source. By assuming the emission lines are broadened
primarily by the virial gas motions in the gravitational potential
of the central object, the BLR size and the line width then give an
estimate of the mass of the central object (Peterson \& Wandel 1999).

Wandel et al. (1999) compiled reverberation-based BLR sizes for
17 Seyfert-1 galaxies and derived a scaling relation between
BLR size and AGN luminosity. Kaspi et al. (2000)\footnote{All
spectra from this monitoring campaign are now publicly available at
http://wise-obs.tau.ac.il/$\sim$shai/PG/.} measured reverberation-based
BLR sizes for 17 high-luminosity AGNs from the Palomar-Green sample
of quasars. Combining these measurements with those of the 17 Seyfert
galaxies, Kaspi et al. (2000)  found that the BLR size scales with AGN
optical luminosity as $R_{\rm BLR}\propto L^{0.70\pm0.03}$. Assuming
that this scaling relation is universal to AGNs of all luminosities
and redshifts, several recent studies have used the relation to
estimate the central masses in large AGN samples using a single-epoch
measurement of the luminosity and the line width (e.g., Wang \& Lu
2001; Woo \& Urry 2002; Grupe \& Mathur 2004). Since the reverberation
studies above were based on the Balmer emission lines, generally broad
H$\beta$, and on the optical AGN luminosity, single-epoch estimates
for objects at redshifts $z\ga 0.6$ have had to rely either on IR
observations (e.g., Shemmer et al. 2004) or on attempts to extend the
optically based size-luminosity relation to UV luminosities and UV
broad emission lines (McLure \& Jarvis 2002; Vestergaard 2002).
While important progress has been made, there are still a number
of potential problems that need to be addressed (e.g., Maoz 2002;
Shemmer et al. 2004; Baskin \& Laor 2005).

In a recent study, Peterson et al. (2004) compiled all available
reverberation mapping data, obtained over the past 15 years,
and analyzed them in a uniform and self-consistent way to improve
the estimates of the black-hole masses. They find that for a given
luminosity, the black-hole mass can be predicted to a precision (random
component of the error) of typically about 30\%. However, there is
also a systematic component of the error, due to the unknown geometry
and kinematics of the BLR, that amounts to a factor of about three,
based on the scatter in the relationship between black hole mass and
bulge velocity dispersion (Onken et al. 2004).  In the present paper
we use the BLR sizes compiled by Peterson et al.  (2004) and study
their relationship with the AGN luminosity measured in X-ray, UV,
and optical continuum bands and in the broad H$\beta$ emission line.
In \S~2 we explain how the data were derived, in \S~3 we present the
relation of the BLR size to the variously defined luminosities, and
in \S~4 we discuss our results. In \S~5 we present a short summary.
Unless otherwise noted all wavelength mentioned in this work are
common rest-frame wavelengths.

\vglue 1cm

\section{Data}
\label{lumi}

Delay times for 35 AGNs with single or multiple data sets were
calculated by Peterson et al (2004). For the current study we use
the Balmer line time delays, as listed in their Table~6, Column
3 (``rest-frame time lags'').  As noted by Kaspi et al. (2000),
the derived BLR size from the first three Balmer lines (H$\alpha$,
H$\beta$, and H$\gamma$) are consistent with each other and averaging
them reduces the scatter in the relation.  We have calculated
the error-weighted mean BLR size derived from these lines where
available. We used only results which were designated by Peterson
et al. (2004) as reliable (see their section 4).  While we do not
necessarily expect each of the Balmer lines to have the same lag,
we find empirically that the precision of the lag measurement for any
one line is sufficiently low that averaging the Balmer line lags for
any particular source reduces the scatter in the radius-luminosity
relationship. We find a similar result when we average multiple
H$\beta$ lag and optical luminosity measurements for a single source.

Note that in their final mass--luminosity relation Peterson et
al. (2004) excluded several data sets for which the full width at half
maximum (FWHM) determination was deemed to be unreliable.  Since FWHM
does not enter the present analysis we have included several of these
data sets if their measured time lag was designated as reliable. We
have  also included an additional data set for NGC\,4051 presented by
Shemmer et al. (2003). The mean BLR size for each data set is listed
in Table~\ref{agndata1}, Column 3. The described analysis below was
done twice -- once using the Balmer-line averaged BLR
size versus the luminosity, and a second time using only the H$\beta$
BLR sizes.  The two analysis sets are discussed in \S~\ref{average}.
In the case where only the H$\beta$ BLR sizes were used, 3 objects
(PG\,0844+349, PG\,1211+143, and NGC4593) were excluded, since their
H$\beta$-based BLR size measurement is considered unreliable.

In the following subsections we explain how the luminosity at each
wavelength band was calculated. We assume a standard cosmology
with $H_{0}=70$ km\,s$^{-1}$\,Mpc$^{-1}$, $\Omega_{M}=0.3$,
and $\Omega_{\Lambda}=0.7$.  The luminosities are tabulated in
Tables~\ref{agndata1} and \ref{agndata2} together with the measured
fluxes (in the observer's frame) from which they were derived.

\begin{deluxetable*}{lcccccc}
\tablecolumns{7}
%\tabletypesize{\scriptsize}
%\rotate
\tablewidth{400pt}
\tablecaption{Optical Observed Fluxes and Luminosities
\label{agndata1}}
\tablehead{
\colhead{Object} &
\colhead{Ref.\tablenotemark{a}} &
\colhead{Time lag\tablenotemark{b}} &
\colhead{$F$(H$\beta$)} &
\colhead{$L$(H$\beta$)} &
\colhead{$f_\lambda$(5100\AA )} &
\colhead{$\lambda L_{\lambda}$(5100\AA )}    \\
\colhead{} &
\colhead{} &
\colhead{} &
\colhead{$10^{-13}$} &
\colhead{$10^{43}$} &
\colhead{$10^{-15}$} &
\colhead{$10^{44}$}  \\
\colhead{} &
\colhead{} &
\colhead{lt-days} &
\colhead{\ergcms} &
\colhead{\ergs} &
\colhead{\ergcmsA} &
\colhead{\ergs}   \\
\colhead{(1)} &
\colhead{(2)} &
\colhead{(3)} &
\colhead{(4)} &
\colhead{(5)} &
\colhead{(6)} &
\colhead{(7)} 
 }
\startdata
Mrk\,335     &  1    & $ 16.8^{+ 4.8}_{- 4.2}$ & $ 7.88\pm 0.36$ & $  0.1328\pm  0.0060$ & $ 7.68\pm 0.53$ & $  0.678\pm  0.047$ \\
            &  1    & $ 12.5^{+ 6.6}_{- 5.5}$ & $ 8.63\pm 0.21$ & $  0.1455\pm  0.0036$ & $ 8.81\pm 0.47$ & $  0.777\pm  0.041$ \\
   &  & { \bf 14.7$^{+ 3.0}_{- 3.0}$}  &  \nodata & { \bf 0.1392$\pm$  0.0048} &  \nodata & { \bf 0.727$\pm$0.044} \\
PG\,0026+129  &  2    & $104.7^{+18.3}_{-18.9}$ & $ 1.27\pm 0.11$ & $  0.8262\pm  0.0698$ & $ 2.69\pm 0.40$ & $10.3\pm 1.5$ \\
PG\,0052+251  &  2    & $ 84.4^{+15.3}_{-13.3}$ & $ 1.80\pm 0.21$ & $ 1.3290\pm  0.1567$ & $ 2.07\pm 0.37$ & $ 9.1\pm 1.6$ \\
Fairall\,9         &  3    & $ 17.4^{+ 3.2}_{- 4.3}$ & $ 7.97\pm 0.37$ & $  0.4494\pm  0.0207$ & $ 5.95\pm 0.66$ & $ 1.79\pm  0.20$ \\
Mrk\,590     &  1    & $ 20.7^{+ 3.5}_{- 2.7}$ & $ 4.41\pm 0.29$ & $  0.0783\pm  0.0052$ & $ 7.89\pm 0.62$ & $  0.736\pm  0.058$ \\
            &  1    & $ 14.0^{+ 8.5}_{- 8.8}$ & $ 2.11\pm 0.71$ & $  0.0375\pm  0.0126$ & $ 5.33\pm 0.56$ & $  0.497\pm  0.053$ \\
            &  1    & $ 29.2^{+ 4.9}_{- 5.0}$ & $ 3.05\pm 0.32$ & $  0.0542\pm  0.0058$ & $ 6.37\pm 0.45$ & $  0.594\pm  0.042$ \\
            &  1    & $ 28.8^{+ 3.6}_{- 4.2}$ & $ 4.44\pm 0.44$ & $  0.0789\pm  0.0079$ & $ 8.43\pm1.30$ & $  0.786\pm  0.122$ \\
    &  & { \bf 23.2$^{+ 7.3}_{- 7.3}$} &  \nodata & { \bf 0.0622$\pm$0.0079} &  \nodata & { \bf 0.653$\pm$0.069} \\
3C\,120       &  1    & $ 38.1^{+21.3}_{-15.3}$ & $ 3.78\pm 0.37$ & $  0.2311\pm  0.0227$ & $ 4.30\pm 0.77$ & $ 1.39\pm  0.25$ \\
Akn\,120      &  1    & $ 47.1^{+ 8.3}_{-12.4}$ & $10.64\pm 0.67$ & $  0.3753\pm  0.0236$ & $10.37\pm 0.46$ & $ 1.927\pm  0.086$ \\
            &  1    & $ 37.1^{+ 4.8}_{- 5.4}$ & $ 7.31\pm 0.83$ & $  0.2580\pm  0.0294$ & $ 7.82\pm 0.83$ & $ 1.45\pm  0.16$ \\
   &  & { \bf 42.1$^{+ 7.1}_{- 7.1}$} &  \nodata & { \bf 0.3167$\pm$0.0265} &  \nodata & { \bf 1.69$\pm$0.12} \\
Mrk\,79       &  1    & $  9.0^{+ 8.3}_{- 7.8}$ & $ 5.38\pm 0.33$ & $  0.0747\pm  0.0046$ & $ 6.96\pm 0.67$ & $  0.502\pm  0.048$ \\
            &  1    & $ 16.1^{+ 6.6}_{- 6.6}$ & $ 5.79\pm 0.48$ & $  0.0803\pm  0.0067$ & $ 8.49\pm 0.86$ & $  0.612\pm  0.062$ \\
            &  1    & $ 16.0^{+ 6.4}_{- 5.8}$ & $ 5.41\pm 0.28$ & $  0.0750\pm  0.0039$ & $ 7.40\pm 0.72$ & $  0.534\pm  0.052$ \\
   &  & { \bf 13.7$^{+ 4.1}_{- 4.1}$} &  \nodata & { \bf   0.0767$\pm$0.0051} &  \nodata & { \bf   0.550$\pm$0.054} \\
PG\,0804+761  &  2    & $162.5^{+31.2}_{-31.2}$ & $ 6.45\pm 0.42$ & $ 1.8020\pm  0.1170$ & $ 5.48\pm1.00$ & $ 8.6\pm 1.6$ \\
PG\,0844+349  &  2    & $ 32.3^{+13.7}_{-13.4}$ & $ 3.02\pm 0.30$ & $  0.3325\pm  0.0330$ & $ 3.71\pm 0.38$ & $ 2.21\pm  0.23$ \\
Mrk\,110      &  1    & $ 24.3^{+ 5.5}_{- 8.3}$ & $ 4.17\pm 0.17$ & $  0.1249\pm  0.0049$ & $ 3.45\pm 0.36$ & $  0.547\pm  0.057$ \\
            &  1    & $ 20.4^{+10.5}_{- 6.3}$ & $ 3.38\pm 0.22$ & $  0.1011\pm  0.0066$ & $ 3.96\pm 0.51$ & $  0.628\pm  0.080$ \\
            &  1    & $ 33.3^{+14.9}_{-10.0}$ & $ 2.99\pm 0.09$ & $  0.0897\pm  0.0025$ & $ 2.64\pm 0.86$ & $  0.42\pm  0.14$ \\
   &  & { \bf  26.0$^{+ 6.6}_{- 6.6}$} &  \nodata & { \bf   0.1052$\pm$0.0047} &  \nodata & { \bf   0.531$\pm$0.091} \\
PG\,0953+414  &  2    & $151.1^{+16.9}_{-21.8}$ & $ 1.60\pm 0.09$ & $ 2.6980\pm  0.1571$ & $ 1.56\pm 0.21$ & $16.6\pm 2.2$ \\
NGC\,3227     &  5,6  & $ 11.4^{+ 4.4}_{- 6.7}$ & $12.53\pm1.95$ & $  0.0044\pm  0.0007$ & $14.41\pm2.46$ & $  0.0256\pm  0.0044$ \\
            &  7    & $  5.4^{+14.1}_{- 8.7}$ & $ 5.10\pm 0.43$ & $  0.0018\pm  0.0002$ & $12.70\pm 0.68$ & $  0.0228\pm  0.0012$ \\
   &  & { \bf   8.4$^{+ 4.2}_{- 4.2}$} &  \nodata & { \bf   0.0031$\pm$0.0004} &  \nodata & { \bf   0.0242$\pm$0.0028} \\
NGC\,3516     &  7,8  & $ 10.9^{+ 4.4}_{- 2.1}$ & $10.82\pm10.23$ & $  0.0216\pm  0.0025$ & $ 7.83\pm2.35$ & $  0.077\pm  0.023$ \\
NGC\,3783     &  9,10 & $ 10.2^{+ 3.3}_{- 2.3}$ & $10.13\pm 0.79$ & $  0.0308\pm  0.0024$ & $11.38\pm 0.95$ & $  0.178\pm  0.015$ \\
NGC\,4051     &  12   & $  5.8^{+ 2.6}_{- 1.8}$ & $ 5.02\pm 0.51$ & $  0.0006\pm  0.0001$ & $13.38\pm 0.92$ & $  0.0086\pm  0.0006$ \\
            &  28   & $  2.8^{+ 1.3}_{- 1.3}$ & $ 3.09\pm 0.26$ & $  0.0004\pm  0.0001$ & $13.86\pm 0.48$ & $  0.0089\pm  0.0003$ \\
  &  & { \bf   4.3$^{+ 2.1}_{- 2.1}$} &  \nodata & { \bf   0.0005$\pm$0.0001} &  \nodata & { \bf   0.0087$\pm$0.0004} \\
NGC\,4151     &  13   & $  3.1^{+ 1.1}_{- 1.0}$ & $72.83\pm4.66$ & $  0.0193\pm  0.0012$ & $81.67\pm4.72$ & $  0.1110\pm  0.0064$ \\
            &  26   & $ 11.2^{+ 2.7}_{- 2.4}$ & $21.6\pm16.2$ & $  0.0057\pm  0.0043$ & $38.31\pm2.81$ & $  0.0519\pm  0.0038$ \\
   &  & { \bf   7.1$^{+ 5.7}_{- 5.7}$} &  \nodata & { \bf   0.0125$\pm$0.0028} &  \nodata & { \bf   0.0815$\pm$0.0051} \\
PG\,1211+143  &  2    & $ 93.2^{+19.7}_{-29.9}$ & $ 5.45\pm 0.67$ & $  0.9727\pm  0.1193$ & $ 5.66\pm 0.92$ & $ 5.57\pm  0.90$ \\
PG\,1226+023  &  2    & $352.4^{+73.3}_{-73.3}$ & $17.22\pm1.40$ & $12.4600\pm 1.0160$ & $21.30\pm2.60$ & $91.1\pm11.1$ \\
PG\,1229+204  &  2    & $ 33.5^{+14.6}_{-12.4}$ & $ 1.98\pm 0.25$ & $  0.2045\pm  0.0258$ & $ 2.15\pm 0.23$ & $ 1.21\pm  0.13$ \\
NGC\,4593     &  6,14 & $  3.2^{+ 5.5}_{- 4.1}$ & $ 3.95\pm1.11$ & $  0.0077\pm  0.0022$ & $12.20\pm3.90$ & $  0.122\pm  0.039$ \\
PG\,1307+085  &  2    & $111.4^{+32.8}_{-43.2}$ & $ 1.92\pm 0.26$ & $ 1.3720\pm  0.1830$ & $ 1.79\pm 0.18$ & $ 7.54\pm  0.76$ \\
IC\,4329A     &  15   & $  1.5^{+ 2.7}_{- 1.8}$ & $ 3.05\pm 0.30$ & $  0.0212\pm  0.0021$ & $ 5.79\pm 0.73$ & $  0.208\pm  0.026$ \\
Mrk\,279      &  16   & $ 16.7^{+ 3.9}_{- 3.9}$ & $ 5.73\pm 0.40$ & $  0.1282\pm  0.0089$ & $ 6.90\pm 0.69$ & $  0.810\pm  0.082$ \\
            &  25   & $ 12.4^{+ 9.5}_{-11.9}$ & $ 2.46\pm 0.43$ & $  0.0550\pm  0.0096$ & $ 5.80\pm1.95$ & $  0.68\pm  0.23$ \\
   &  & { \bf  14.6$^{+ 3.0}_{- 3.0}$} &  \nodata & { \bf   0.0916$\pm$0.0093} &  \nodata & { \bf   0.75$\pm$0.16} \\
PG\,1411+442  &  2    & $101.6^{+31.0}_{-28.1}$ & $ 3.33\pm 0.18$ & $  0.6829\pm  0.0371$ & $ 3.71\pm 0.32$ & $ 4.22\pm  0.36$ \\
NGC\,5548     &  17   & $ 19.7^{+ 1.5}_{- 1.5}$ & $ 8.62\pm 0.85$ & $  0.0609\pm  0.0060$ & $ 9.92\pm1.26$ & $  0.362\pm  0.046$ \\
            &  17   & $ 18.6^{+ 2.1}_{- 2.3}$ & $ 5.98\pm1.17$ & $  0.0422\pm  0.0083$ & $ 7.25\pm1.00$ & $  0.26\pm  0.037$ \\
            &  17   & $ 15.9^{+ 2.9}_{- 2.5}$ & $ 7.46\pm 0.81$ & $  0.0527\pm  0.0057$ & $ 9.40\pm 0.93$ & $  0.343\pm  0.034$ \\
            &  17   & $ 11.0^{+ 1.9}_{- 2.0}$ & $ 4.96\pm1.44$ & $  0.0350\pm  0.0102$ & $ 6.72\pm1.17$ & $  0.246\pm  0.043$ \\
            &  17   & $ 13.0^{+ 1.6}_{- 1.4}$ & $ 7.93\pm 0.53$ & $  0.0560\pm  0.0037$ & $ 9.06\pm 0.86$ & $  0.331\pm  0.032$ \\
            &  17   & $ 13.4^{+ 3.8}_{- 4.3}$ & $ 7.58\pm 0.94$ & $  0.0535\pm  0.0066$ & $ 9.76\pm1.10$ & $  0.356\pm  0.040$ \\
            &  17   & $ 21.7^{+ 2.6}_{- 2.6}$ & $ 9.27\pm 0.70$ & $  0.0654\pm  0.0050$ & $12.09\pm1.00$ & $  0.442\pm  0.037$ \\
            &  17   & $ 16.4^{+ 1.2}_{- 1.1}$ & $ 7.95\pm 0.87$ & $  0.0561\pm  0.0061$ & $10.56\pm1.64$ & $  0.386\pm  0.060$ \\
            &  17   & $ 17.5^{+ 2.0}_{- 1.6}$ & $ 7.41\pm 0.95$ & $  0.0523\pm  0.0067$ & $ 8.12\pm 0.91$ & $  0.297\pm  0.033$ \\
            &  17   & $ 26.5^{+ 4.3}_{- 2.2}$ & $10.27\pm1.04$ & $  0.0725\pm  0.0073$ & $13.47\pm1.45$ & $  0.492\pm  0.053$ \\
            &  17   & $ 24.8^{+ 3.2}_{- 3.0}$ & $ 9.34\pm 0.61$ & $  0.0659\pm  0.0043$ & $11.83\pm1.82$ & $  0.432\pm  0.066$ \\
            &  17   & $  6.5^{+ 5.7}_{- 3.7}$ & $ 6.27\pm1.22$ & $  0.0443\pm  0.0086$ & $ 6.98\pm1.20$ & $  0.255\pm  0.044$ 
\enddata
\end{deluxetable*}

\setcounter{table}{0}

\begin{deluxetable*}{lcccccc}
\tablecolumns{7}
%\tabletypesize{\scriptsize}
%\rotate
\tablewidth{400pt}
\tablecaption{--- $Continued$}
\tablehead{
\colhead{Object} &
\colhead{Ref.\tablenotemark{a}} &
\colhead{Time lag\tablenotemark{b}} &
\colhead{$F$(H$\beta$)} &
\colhead{$L$(H$\beta$)} &
\colhead{$f_\lambda$(5100\AA )} &
\colhead{$\lambda L_{\lambda}$(5100\AA )}    \\
\colhead{} &
\colhead{} &
\colhead{} &
\colhead{$10^{-13}$} &
\colhead{$10^{43}$} &
\colhead{$10^{-15}$} &
\colhead{$10^{44}$}  \\
\colhead{} &
\colhead{} &
\colhead{lt-days} &
\colhead{\ergcms} &
\colhead{\ergs} &
\colhead{\ergcmsA} &
\colhead{\ergs}   \\
\colhead{(1)} &
\colhead{(2)} &
\colhead{(3)} &
\colhead{(4)} &
\colhead{(5)} &
\colhead{(6)} &
\colhead{(7)} 
 }
\startdata
            &  17   & $ 14.3^{+ 5.9}_{- 7.3}$ & $ 5.26\pm1.12$ & $  0.0372\pm  0.0079$ & $ 7.03\pm 0.86$ & $  0.257\pm  0.032$ \\
            &  27   & $  7.6^{+ 4.8}_{- 4.0}$ & $ 7.10\pm 0.60$ & $  0.0501\pm  0.0042$ & $ 5.78\pm 0.62$ & $  0.211\pm  0.023$ \\
   &  & { \bf  16.2$^{+ 5.9}_{- 5.9}$} &  \nodata & { \bf   0.0532$\pm$0.0065} &  \nodata & { \bf   0.334$\pm$0.041} \\
PG\,1426+015  &  2    & $ 84.7^{+21.4}_{-24.4}$ & $ 3.09\pm 0.28$ & $  0.6302\pm  0.0581$ & $ 4.62\pm 0.71$ & $ 5.21\pm  0.80$ \\
Mrk\,817      &  1    & $ 19.0^{+ 3.9}_{- 3.7}$ & $ 4.73\pm 0.26$ & $  0.1100\pm  0.0061$ & $ 6.10\pm 0.83$ & $  0.75\pm  0.10$ \\
            &  1    & $ 15.3^{+ 3.7}_{- 3.5}$ & $ 4.00\pm 0.40$ & $  0.0930\pm  0.0092$ & $ 5.00\pm 0.49$ & $  0.61\pm  0.06$ \\
            &  1    & $ 33.6^{+ 6.5}_{- 7.6}$ & $ 3.38\pm 0.19$ & $  0.0785\pm  0.0045$ & $ 5.01\pm 0.27$ & $  0.612\pm  0.033$ \\
   &  & { \bf  22.6$^{+ 9.7}_{- 9.7}$} &  \nodata & { \bf   0.0938$\pm$0.0066} &  \nodata & { \bf   0.656$\pm$0.065} \\
PG\,1613+658  &  2    & $ 40.1^{+15.0}_{-15.2}$ & $ 2.02\pm 0.13$ & $  0.9496\pm  0.0632$ & $ 3.49\pm 0.43$ & $ 9.5\pm 1.2$ \\
PG\,1617+175  &  2    & $ 86.8^{+16.0}_{-20.2}$ & $ 1.40\pm 0.17$ & $  0.5123\pm  0.0636$ & $ 1.44\pm 0.25$ & $ 3.00\pm  0.52$ \\
PG\,1700+518  &  2    & $251.8^{+45.9}_{-38.8}$ & $ 1.89\pm 0.10$ & $ 5.5520\pm  0.2882$ & $ 2.20\pm 0.15$ & $42.34\pm 2.89$ \\
3C\,390.3     &  21   & $ 23.6^{+ 6.2}_{- 6.7}$ & $ 2.12\pm 0.17$ & $  0.1950\pm  0.0153$ & $ 1.73\pm 0.28$ & $  0.87\pm  0.14$ \\
Mrk\,509      &  1    & $ 79.6^{+ 6.1}_{- 5.4}$ & $12.30\pm 0.06$ & $  0.3985\pm  0.0020$ & $10.92\pm1.99$ & $ 1.88\pm  0.34$ \\
PG\,2130+099  &  2    & $177.1^{+19.9}_{-12.7}$ & $ 5.12\pm 0.46$ & $  0.5558\pm  0.0498$ & $ 4.84\pm 0.45$ & $ 2.85\pm  0.26$ \\
NGC\,7469     &  23   & $  4.5^{+ 0.6}_{- 0.7}$ & $ 7.80\pm 0.35$ & $  0.0577\pm  0.0026$ & $13.57\pm 0.61$ & $  0.521\pm  0.023$
\enddata
\tablecomments{Bold font lines give average values for objects with
multiple data sets. The mean time lag is computed as a simple mean
and its uncertainty is the rms of the individual measurements. The
mean luminosity is computed as a simple mean and its uncertainty is
the simple mean of the individual uncertainties (which are derived
from the rms of each light curve).}
\tablenotetext{a}{Reference numbers are as in Table 1 of Peterson et al. (2004)
and are given here to facilitate the identification of the data sets;
reference 28 is Shemmer et al. (2003).} 
\tablenotetext{b}{These are rest-frame time lags. For objects in which
more than one Balmer-line time lag was measured we list the averaged
rest-frame time lag.
}
\end{deluxetable*}

\begin{deluxetable*}{lcccccccc}
\tablecolumns{9}
%\tabletypesize{\scriptsize}
%\tabletypesize{\tiny}
%\rotate
\tablewidth{470pt}
\tablecaption{Object Characteristics
\label{agndata2}}
\tablehead{
\colhead{Object} &
\colhead{Redshift} &
\colhead{$E(B-V)$}  &
\colhead{$f_\lambda$(1450\AA )} &
\colhead{$\lambda L_{\lambda}$(1450\AA )} &
\colhead{$F$(2--10 keV)\tablenotemark{a}} &
\colhead{$L$(2--10 keV)\tablenotemark{b}} &
\colhead{Ref.} &
\colhead{$F$(H$\beta_{\rm narrow}$)} \\
\colhead{} &
\colhead{} &
\colhead{} &
\colhead{$10^{-14}$} &
\colhead{$10^{44}$} &
\colhead{$10^{-12}$} &
\colhead{$10^{43}$} & 
\colhead{}  &
\colhead{$10^{-13}$} \\
\colhead{} &
\colhead{} &
\colhead{} &
\colhead{\ergcmsA} &
\colhead{\ergs} &
\colhead{\ergcms} &
\colhead{\ergs}  &   
\colhead{}   &
\colhead{\ergcms} \\
\colhead{(1)} &
\colhead{(2)} &
\colhead{(3)} &
\colhead{(4)} &
\colhead{(5)} &
\colhead{(6)} &
\colhead{(7)} &
\colhead{(8)} &
\colhead{(9)} 
}
\startdata

Mrk\,335     &  0.02578 &  0.035 & $ 7.2\pm 1.0$ & $ 2.12\pm  0.30$ & $ 9.43\pm 0.10$ & $ 1.430\pm  0.015$ & 1  & 0.410 \\
PG\,0026+129  &  0.14200 &  0.071 & $ 1.5\pm 0.5$ & $22.36\pm 7.45$ & $ 6.33\pm 0.13$ & $34.16\pm  0.70$ & 3  & 0.063 \\
PG\,0052+251  &  0.15500 &  0.047 & $ 1.1\pm 0.2$ & $16.88\pm 3.07$ & $10.09\pm 0.17$ & $66.0\pm 1.1$ & 3  & 0.091 \\
Fairall\,9  &  0.04702 &  0.027 & $ 3.5\pm1.0$ & $ 3.38\pm  0.97$ & $19.06\pm1.00$ & $ 9.92\pm  0.52$ & 2  & 0.283 \\
Mrk\,590     &  0.02638 &  0.037 & $ 2.4\pm1.4$ & $  0.75\pm  0.44$ & $27.00\pm2.70$ & $ 4.29\pm  0.43$ & 7  & 0.121 \\
3C\,120       &  0.03301 &  0.297 & $ 1.2\pm1.2$ & $ 4.45\pm 4.23$ & $44.28\pm2.22$ & $11.13\pm  0.56$ & 2  & 0.151 \\
Akn\,120      &  0.03230 &  0.128 & $ 3.5\pm 0.5$ & $ 3.34\pm  0.48$ & $27.34\pm2.73$ & $ 6.57\pm  0.66$ & 5  & 0.364 \\
Mrk\,79       &  0.02219 &  0.071 & $ 2.0\pm1.0$ & $  0.57\pm  0.29$ & $25.96\pm2.60$ & $ 2.90\pm  0.29$ & 7  & 0.221 \\
PG\,0804+761  &  0.10000 &  0.035 & $ 7.0\pm1.0$ & $36.56\pm 5.22$ & $ 8.34\pm 0.07$ & $21.13\pm  0.18$ & 1  & 0. \\
PG\,0844+349  &  0.06400 &  0.037 & $ 1.7\pm 0.5$ & $ 3.41\pm 1.00$ & $ 2.37\pm 0.02$ & $ 2.34\pm  0.020$ & 1  & 0. \\
Mrk\,110      &  0.03529 &  0.013 & $ 1.0\pm 0.5$ & $  0.48\pm  0.24$ & $26.64\pm 0.19$ & $ 7.676\pm  0.055$ & 1  & 0.148 \\
PG\,0953+414  &  0.23410 &  0.013 & $ 2.0\pm1.0$ & $64.3\pm32.1$ & $ 2.58\pm 0.04$ & $42.22\pm  0.65$ & 2  & 0.534 \\
NGC\,3227     &  0.00386 &  0.023 & $  0.04\pm 0.01$ & $  0.0003\pm  0.0001$ & $25.52\pm1.30$ & $  0.0839\pm  0.0043$ & 2  & 0.558 \\
NGC\,3516     &  0.00884 &  0.042 & $ 3.0\pm1.5$ & $  0.106\pm  0.053$ & $73.68\pm3.70$ & $ 1.280\pm  0.064$ & 2  & 0. \\
NGC\,3783     &  0.00973 &  0.119 & $ 4.8\pm1.0$ & $  0.373\pm  0.078$ & $51.49\pm2.55$ & $ 1.085\pm  0.054$ & 2  & 0.656 \\
NGC\,4051     &  0.00234 &  0.013 & $ 1.2\pm 0.2$ & $  0.0023\pm  0.0004$ & $21.21\pm1.10$ & $  0.026\pm  0.0013$ & 2  & 0. \\
NGC\,4151     &  0.00332 &  0.028 & $ 35.\pm 15.$ & $  0.154\pm  0.066$ & $201.\pm10.$ & $  0.490\pm  0.025$ & 2  & 9.912 \\
PG\,1211+143  &  0.08090 &  0.035 & $ 3.0\pm1.0$ & $ 9.8\pm 3.3$ & $ 2.82\pm 0.04$ & $ 4.557\pm  0.065$ & 1  & 0. \\
PG\,1226+023  &  0.15834 &  0.021 & $15.0\pm4.0$ & $  200.6\pm53.5$ & $90.33\pm20.8$ & $  619.\pm  142.$ & 3  & 0. \\
PG\,1229+204  &  0.06301 &  0.027 & $ 2.0\pm1.0$ & $ 3.6\pm 1.8$ & $ 2.54\pm 0.16$ & $ 2.43\pm  0.15$ & 3  & 0.152 \\
NGC\,4593     &  0.00900 &  0.025 & $ 2.0\pm1.0$ & $  0.064\pm  0.032$ & $31.89\pm1.60$ & $  0.575\pm  0.029$ & 2  & 0.217 \\
PG\,1307+085  &  0.15500 &  0.034 & $  0.8\pm 0.2$ & $11.2\pm 2.8$ & $ 3.65\pm 0.35$ & $23.9\pm 2.3$ & 6  & 0.155 \\
IC\,4329A     &  0.01605 &  0.059 &      \nodata   &      \nodata        & $77.46\pm3.90$ & $ 4.49\pm  0.23$ & 2  & 0.253 \\
Mrk\,279      &  0.03045 &  0.016 & $  0.6\pm 0.5$ & $  0.22\pm  0.18$ & $25.92\pm2.59$ & $ 5.52\pm  0.55$ & 5  & 0.325 \\
PG\,1411+442  &  0.08960 &  0.008 & $ 1.4\pm 0.4$ & $ 4.70\pm 1.34$ & $  0.78\pm 0.09$ & $ 1.57\pm  0.18$ & 1  & 0.140 \\
NGC\,5548     &  0.01717 &  0.020 & $ 3.0\pm1.0$ & $  0.34\pm  0.11$ & $46.05\pm2.30$ & $ 3.06\pm  0.15$ & 2  & 0.614 \\
PG\,1426+015  &  0.08647 &  0.032 & $ 7.5\pm2.0$ & $27.82\pm 7.42$ & $ 5.25\pm 0.25$ & $ 9.77\pm  0.47$ & 3  & 0.072 \\
Mrk\,817      &  0.03145 &  0.007 & $ 1.5\pm1.0$ & $  0.54\pm  0.36$ &      \nodata   &      \nodata  & \nodata  & 0.078 \\
PG\,1613+658  &  0.12900 &  0.027 & $ 3.5\pm1.0$ & $30.52\pm 8.72$ & $ 6.14\pm 0.09$ & $26.90\pm  0.39$ & 3  & 0.041 \\
PG\,1617+175  &  0.11244 &  0.042 & $  0.75\pm 0.35$ & $ 5.35\pm 2.50$ &      \nodata   &      \nodata  & \nodata  & 0. \\
PG\,1700+518  &  0.29200 &  0.035 & $  0.6\pm 0.4$ & $39.4\pm26.3$ &      \nodata   &      \nodata  & \nodata  & 0. \\
3C\,390.3     &  0.05610 &  0.071 & $  0.6\pm 0.2$ & $ 1.17\pm  0.39$ & $22.80\pm5.00$ & $17.11\pm 3.75$ & 4  & 0.172 \\
Mrk\,509      &  0.03440 &  0.057 & $ 5.5\pm1.0$ & $ 3.48\pm  0.63$ & $43.10\pm2.10$ & $11.79\pm  0.57$ & 2  & 0.777 \\
PG\,2130+099  &  0.06298 &  0.044 & $ 2.2\pm 0.2$ & $ 4.50\pm  0.41$ & $ 4.80\pm 0.12$ & $ 4.59\pm  0.11$ & 3  & 0.385 \\
NGC\,7469     &  0.01632 &  0.069 & $ 3.0\pm 0.5$ & $  0.451\pm  0.075$ & $27.64\pm3.40$ & $ 1.66\pm  0.20$ & 8  & 1.357 \\
\enddata
\tablenotetext{a}{Observed X-ray flux in the observed energy band 2--10 keV.}
\tablenotetext{b}{X-ray luminosity in the rest-frame energy band 2--10 keV.}
\tablerefs{{\sc For column 8:} 1 - George et al. (2000);
2 - George et al. (1998); 3 - Lawson \& Turner (1997); 4 - Leighly et
al. (1997); 5 - The Tartarus Database (http://tartarus.gsfc.nasa.gov/);
6 - Williams et al. (1992); 7 - Turner \& Pounds (1989); 8 - Nandra
et al. (1998).
}
\end{deluxetable*}

\subsection{General Considerations for the Multiwavelength Datasets}

We derived optical luminosities directly from the data sets from which
the time lags were measured. Hence, the measurement of the time lag
and the optical luminosity is simultaneous. However, the UV and X-ray
fluxes were measured simultaneously with the BLR size for only a few
AGNs. In these objects, we used these simultaneous UV and X-ray data
in the analysis below (UV -- NGC\,3783, NGC\,5548, NGC\,4151, 3C390.3,
NGC7469, Fairall\,9; X-ray -- NGC4151, NGC7469). Luminosities were not
corrected for host galaxy contribution. To subtract this contribution,
a knowledge of the spatial distribution of the host-galaxy luminosity
is needed, combined with the slit width and seeing in which the
observation was done. This task is beyond the scope of the current
paper.  However, we note that this correction would be negligible
for the quasars in the sample ($\lambda L_\lambda$(5100\AA )$ \ga
10^{44}$ \ergs ) and it is likely small for Seyferts as well (e.g.,
Peterson et al. 1995; Kaspi et al. 1996). Since there might be a
systematic effect of greater host-galaxy luminosity contribution
in lower luminosity objects, such a correction will tend to flatten
the  BLR-size -- luminosity relation causing the power-law slope to
be smaller.

Recent studies (e.g., Gaskell et al. 2004; Hopkins et al. 2004)
indicate AGNs have reddening in addition to the Galactic reddening,
probably due to gas in their host galaxy or the intergalactic
medium. This is affecting the measured luminosity of the AGNs, though
it is difficult to determine the intrinsic reddening of individual
objects. Hence, we do not attempt to correct for such reddening in
the study presented here.

AGNs vary with differing amplitudes and timescales at different
wavelengths. Specifically, amplitudes become larger and timescales
shorter as one progresses from the rest-frame optical, through the
UV, to X-rays (e.g., Edelson et al. 1996). Therefore, another basic
limitation of the comparison we make of the size-luminosity relation
for luminosities in different bands, beyond the fact that the actual
sampling window in each band is very different (e.g., robust median
luminosities in the optical, based on tens of epochs, compared
with few- or single-epoch-based luminosities in UV and X-rays),
is the different statistical properties of the variability within each
band.

Table~\ref{agndata1} lists all multiple data sets. One line per object
(in bold font) gives the mean of all the data sets for this object
(cases in which there is only one data set appear with a normal font).
This mean will be further discussed in \S~\ref{average}.

\subsection{Optical luminosities}

Optical luminosities were calculated from the mean 5100~\AA\
flux density of each data set. These luminosities are the same as
the ones in Peterson et al. (2004; see their \S~8), but with an
improved correction for Galactic absorption. We used the $E(B-V)$
values listed in the NASA/IPAC Extragalactic Database (NED) which
are taken from Schlegel et al. (1998) and the extinction curve of
Cardelli et al. (1989), adjusted to $A_V/E(B-V)=3.1$.  For most
objects the difference in luminosities between the current work and
Peterson et al. (2004) is less than 1\%, while for a few it reaches
up to 7\%.  The measured 5100~\AA\ flux densities and luminosities
are listed in Table~\ref{agndata1}, Columns 6 and 7, respectively.
The luminosity uncertainty is determined as the rms of the light
curve of each data set.

We have computed the H$\beta$ luminosity from the average
H$\beta$ flux of each data set, correcting the fluxes for Galactic
absorption as above. The H$\beta$ flux and luminosity are listed in
Table~\ref{agndata1}, Columns 4 and 5, respectively. We have also
estimated the H$\beta$ luminosity with the narrow component removed
from the line profile (these narrow-line corrected H$\beta$ fluxes and
luminosities do not appear in Table~\ref{agndata1}). The contribution
of the narrow component of H$\beta$ was calculated using the
H$\beta$(narrow component)/[\ion{O}{3}]$\lambda$5007 line ratio from
Marziani et al. (2003), multiplied by the [\ion{O}{3}]$\lambda$5007
line flux we measured in the data sets in hand.  We used our own
estimates for the narrow component of H$\beta$ for: a) AGNs which
are not in the  Marziani et al. sample (NGC\,3227, NGC\,4151);
b) AGNs which were observed at a low-state and for which we have
a particularly good estimate of the narrow component of H$\beta$
(Mrk\,79, PG\,1229+204, NGC\,5548, PG\,1426+015). We also did not
use the Marziani et al. estimates for five objects (PG\,0804+761,
PG\,0844+349, PG\,1211+143, PG\,1226+023, PG\,1617+175) for which we
think their [\ion{O}{3}]$\lambda$5007 line fluxes are overestimated
due to blending with \ion{Fe}{2}$\lambda$5018. The H$\beta$
narrow component line flux we use in our analysis is listed in
Table~\ref{agndata2}, Column 9.

\vglue 1cm

\subsection{UV luminosity}
\vglue -0.1cm

For the UV luminosity, we have used both the 1350~\AA\ (used by
Vestergaard 2002) and 1450~\AA\ (used by Shemmer et al. 2004)
wavelength ranges.  UV fluxes at these rest frame wavelengths were
measured, for most objects, using FOS and STIS spectra taken from the
{\it Hubble Space Telescope (HST)} archive. In cases where multiple
observations exist, we have used the averaged flux density, except
for the few cases where the UV data were obtained simultaneously with
the optical time lag measurements -- see \S~\ref{lumi}. In several
cases (Mrk\,110, PG\,1229+204, PG\,1426+015, Mrk\,817, PG\,1613+658)
there were no observations at 1350~\AA\ or 1450~\AA\ but spectra
at about 1200~\AA\ could be extrapolated (using a straight line)
in order to estimate the 1350~\AA\ and 1450~\AA\ flux densities. For
several objects without {\it HST} UV spectra we used archival {\it
International Ultraviolet Explorer (IUE)} spectra to measure the
1350~\AA\ and 1450~\AA\ flux densities. These objects are 3C\,120,
Mrk\,590, Mrk\,79, and PG\,1617+175. For one object, IC\,4329A,
no useful UV data were found. The flux densities were corrected
for Galactic absorption, as described above for the 5100~\AA\
flux density. The measured 1450~\AA\ flux densities and calculated
luminosities are listed in Table~\ref{agndata2} in Columns 4 and 5,
respectively.
\vglue -0.8cm

\subsection{X-ray luminosity}

Most of the objects with reverberation mapping data also have
X-ray data in the literature and in archives.  We searched the
literature for the 2--10 keV fluxes of each object (see references
in Table~\ref{agndata2}, column 8).  For 3 objects (PG\,1700+519,
Mrk\,817, and PG\,1617+175), no useful X-ray data were found. The
fluxes were corrected for Galactic absorption using the PIMMS tool
(Mukai 1993) and the published power-law slopes and Galactic column
densities.  The observed 2--10 keV band was K-corrected to yield
the rest-frame 2--10 keV luminosity. We note, however, that because
of the small Galactic absorption of most objects and their small
redshifts, these corrections are small (of order of a few percent
and hence much smaller than the known X-ray variability). Also,
any intrinsic absorption has a negligible effect in this hard X-ray
energy range. The measured fluxes in the observed 2--10 keV band
and the calculated luminosities in the rest-frame 2--10 keV band are
listed in Table~\ref{agndata2} Columns 6 and 7, respectively.

\section{BLR Size -- Luminosity Relations}
\label{average}

\begin{figure*}
\centerline{\includegraphics[width=17.0cm]{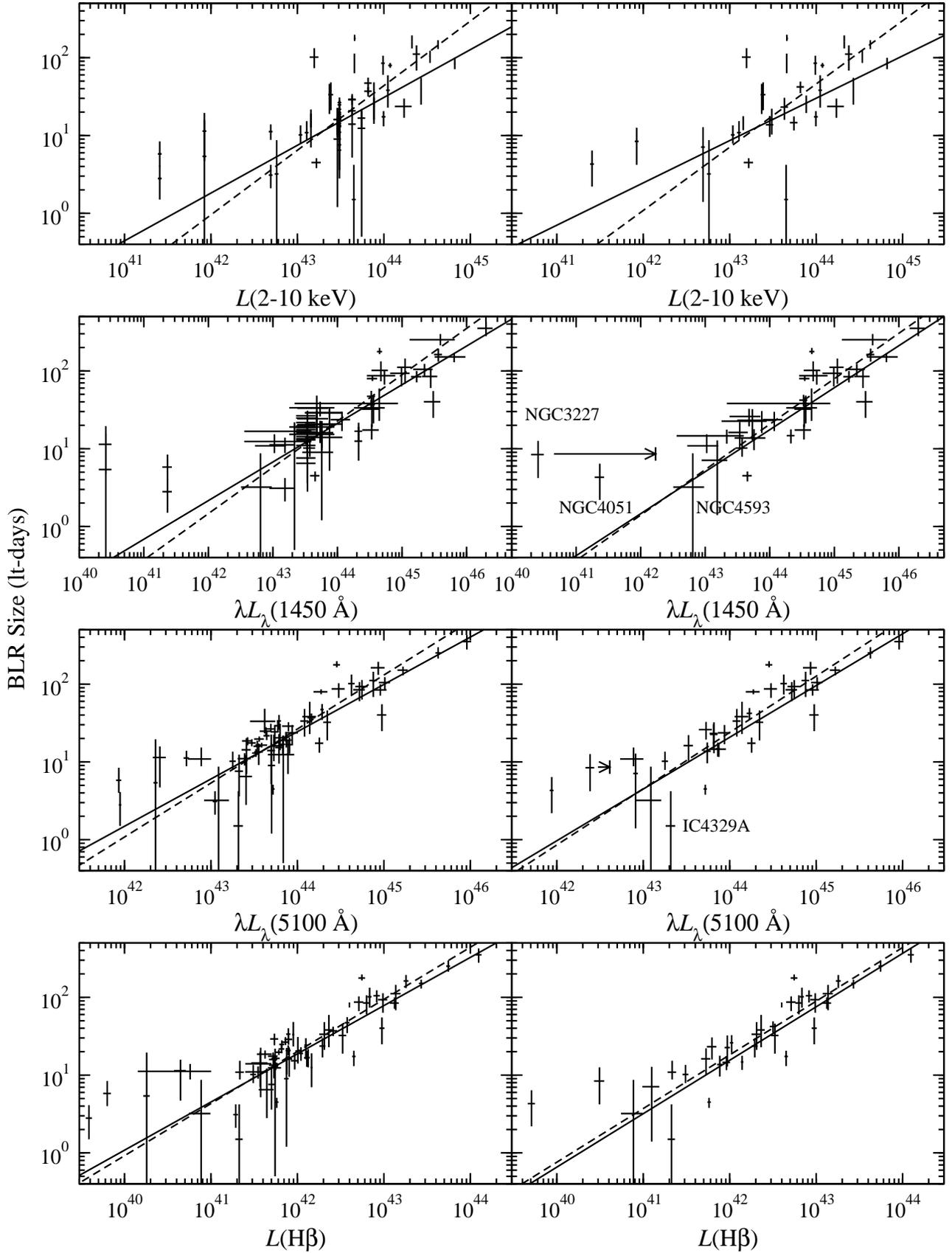}}
\caption{\scriptsize Balmer-line averaged BLR size plotted versus
various luminosities given in \ergs . The left side panels show
all data sets and the right side panels show one averaged data
set per object.  The data are listed in Tables~\ref{agndata1}
and \ref{agndata2}. For each panel two fits (excluding the
two lowest luminosity objects) are shown: solid line -- FITEXY and
dashed line - BCES. Fits parameters are given in the upper part of
Table~\ref{results}. Right pointing arrows show the shift in location
of the data points of NGC\,3227 when the Crenshaw et al. (2001)
extinction correction is applied.
\label{mainfigure} }
\end{figure*}

The relations between the BLR size and the various luminosities
described above are plotted in Figure~\ref{mainfigure}. The power-law
relation between the BLR size and $\lambda L_\lambda$(5100 \AA ) is
clearly confirmed by our analysis. Such a relation is also apparent
when UV, X-ray, and H$\beta$ luminosities are used.  Two obvious
outliers, which do not fit the general trends in any of the wavelength
bands, are NGC\,3227 and NGC\,4051. These are the two-least luminous
objects in all wavelength bands. This behavior could be due to major
reddening or obscuration in these two objects, which decreases their
measured luminosity (e.g., NGC\,3227 - Crenshaw et al. 2001; NGC\,4051
- Kaspi et al. 2004).  For NGC\,3227, the extinction correction
suggested by Crenshaw et al. (2001) increases the UV luminosity
(at 1450 \AA ) by a factor of $\sim 66$ and the optical luminosity
by a factor of $\sim 2$ (see Figure~\ref{mainfigure}). Although this
would improve the agreement of this data point with the general trend,
it would still remain an outlier. An alternative explanation can be
that the BLR-size -- luminosity relation found at high luminosities
breaks down at low luminosities. We will exclude these two objects from
our fits to the BLR-size -- luminosity relation, noting the derived
relation applies to AGNs with optical luminosity in the range $
10^{43} < \lambda L_{\lambda}(5100{\rm \AA}) < 10^{46}  $ \ergs .

We have used two methods to calculate the relations between the BLR
size and the various luminosities:
\begin{description}
\item{1. The linear regression method of Press et al. (1992), in
which a straight-line is fitted to the data with errors in both
coordinates (known as FITEXY). This method is based on an iterative
process to minimize $\chi^2$. We follow Tremaine et al. (2002),
who account for the intrinsic scatter in the relation by increasing
the uncertainties\footnote{In the present work we increase the
uncertainty of the time lags by a certain percent of the measured
value.} until obtaining $\chi^2$ per degree of freedom equal to 1
(see below in Table~\ref{results} column 5).}
\item{2. The bivariate correlated errors and intrinsic scatter (BCES)
regression method of Akritas \& Bershady (1996). This method takes into
account the uncertainty in both coordinates as well as the intrinsic
scatter around a straight line. In our analysis, we use the BCES
bisector result which is the mean of the two fits: one when fitting
Y as dependent on X, and the other when fitting X as dependent on Y.}
\end{description}
Tremaine et al. (2002) compare the two methods using Monte-Carlo
simulations and concluded that the FITEXY is superior to the BCES
method. However, in their implementation for the BCES method Tremaine
et al. (2002) did not use the BCES bisector value.  They only used
the fit of Y as dependent on X. This might have distorted their
conclusion when comparing the two methods. Thus in the analysis below
we use both methods.  One advantage of the BCES method over the simple
FITEXY method is its ability to account for real intrinsic scatter, if
it exists in the data. However, the available code of the BCES method
does not give any quantification of this scatter. The implementation
of the FITEXY method that we use (of Tremaine et al. 2002)
also considers intrinsic scatter and allows us to quantify it.
Such intrinsic scatter was assumed in previous investigations of
the BLR-size -- luminosity relation (e.g., Vestergaard 2002), but
not quantified.

The two fitting methods do not account for asymmetric uncertainties
(e.g., as we have for the uncertainties of the BLR sizes).  In order
to account for the asymmetric uncertainties, we used in our fits the
uncertainty value in the direction of the fitted line. This typically
required a few iterations to converge completely.

The correlation analysis below is done in two ways: (1) by using all
data sets that are available, where in case of multiple data sets for
the same object we treat each data set as independent; (2) by computing
the mean BLR size and mean luminosity of all the multiple
data sets of the same object, and using only that mean for each object.

In Table~\ref{results}, we summarize the results from the different
fits we carried out for the relation of the various luminosities
with the BLR size. For each of the luminosities measured, two lines
are given: the first line lists the results when using all data sets
and the second line lists the results when using the averaged data
set per object.  The number of data sets used in the fit is listed
in Column 2. The relations are given in the form:
\begin{equation}
\frac{R_{BLR}}{10 {\rm \ lt\,days}} = A L^{B}  
\label{eqfit} 
\end{equation}
where $L$ is the luminosity normalized as given in column 1 of
Table~\ref{results}. The parameters A and B are given in columns 3 and
4 for the FITEXY method and in columns 6 and 7 for the BCES method.
In column 5 we list the intrinsic scatter, in percent of the value,
found when using  the FITEXY method. Also, in columns 8 and 9 we
list the Pearson and Spearman correlation coefficients, respectively.

As explained in \S~\ref{lumi}, the analysis was done twice -- once
using the mean Balmer-line BLR size versus the luminosity
(results are presented in the upper part of Table~\ref{results}),
and a second time using only the H$\beta$ BLR sizes (lower part of
Table~\ref{results}).  The results from the two analysis sets are
consistent and no significant differences are detected. The correlations
coefficients (columns 8 and 9 of Table~\ref{results}) are somewhat higher
when using the mean Balmer-line BLR size and this may indicate that
using these lags reduces the scatter in the relation with the
luminosity. Thus, in the analysis below we refer only to the upper
part of Table~\ref{results} which uses the mean Balmer-lines time lags.

The results of Table~\ref{results} indicate that there is a systematic
difference between the two fitting methods. The BCES method slope
parameter (B) is generally larger than the one found with the FITEXY
fit, and also the constant parameter (A) is generally larger. This
illustrates the important issue of how the regression method may
influence the correlation results; using different regression methods
yields different regression slopes.  We could not find what is the
main cause for the differences and we attribute them to the different
algorithms, strategies, and statistical concepts each of the two
methods uses. The real uncertainty of the slope measurement should
be the range between the different methods and not necessarily the
formal error given by each method. Furthermore, the regression methods
are sensitive also to outlier points. We find that the BCES method
is much more sensitive to the inclusion or exclusion of the two
lower-luminosity points (NGC\,3227 and NGC\,4051) compared to FITEXY.

In its present implementation the FITEXY method establishes
that intrinsic scatter does exist in the BLR size -- luminosity
relation. Only when we artificially increase the formal measured
uncertainty in the BLR size do we find a $\chi^2$ of unity. We
find that the intrinsic scatter in the relation is $\sim 40$\%
(Table~\ref{results} Column 5). We note that this intrinsic scatter
can be caused by a real scatter in the physical BLR size -- luminosity
relation but also by some systematic effects such as intrinsic
reddening, contributions by the host galaxy, or effects of variability
due to non-contemporaneous observations. There is currently no way
to distinguish between the different contributors to this scatter.

\begin{deluxetable*}{lcccccccc}
\tablecolumns{9}
%\tabletypesize{\scriptsize}
%\rotate
\tablewidth{450pt}
\tablecaption{BLR size --- Luminosity relation\tablenotemark{a}
\label{results}}
\tablehead{
\colhead{Luminosity} &
\colhead{N} &
\colhead{A$_{\rm FITEXY}$} &
\colhead{B$_{\rm FITEXY}$} &
\colhead{Scatter\tablenotemark{b}} &
\colhead{A$_{\rm BCES}$} &
\colhead{B$_{\rm BCES}$} &
\colhead{Pearson} &
\colhead{Spearman} \\
\colhead{(1)} &
\colhead{(2)} &
\colhead{(3)} &
\colhead{(4)} &
\colhead{(5)} &
\colhead{(6)} &
\colhead{(7)} &
\colhead{(8)} &
\colhead{(9)} 
}
\startdata
\cutinhead{Using mean Balmer-lines time lag}
%
% \colhead{Luminosity}                 & N  &            A_ FITEXY   &   B_FITEXY      & EPS &     A_BCES             &     B_BCES      & Pears & Spear \\
$L$(H$\beta$)/10$^{43}$                & 59 & $7.83^{+0.87}_{-0.78}$ & $0.619\pm0.045$ & 38  & $9.42^{+0.93}_{-0.85}$ & $0.670\pm0.050$ & 0.885 & 0.860 \\
                                       & 33 & $7.55^{+0.90}_{-0.81}$ & $0.687\pm0.058$ & 40  & $8.91^{+0.92}_{-0.83}$ & $0.690\pm0.068$ & 0.916 & 0.933 \\  
$L$(H$\beta$)/10$^{43}$ no narrow      & 59 & $8.14^{+0.91}_{-0.82}$ & $0.618\pm0.044$ & 38  & $9.68^{+0.99}_{-0.90}$ & $0.659\pm0.052$ & 0.879 & 0.872 \\ 
                                       & 33 & $7.87^{+0.89}_{-0.80}$ & $0.683\pm0.055$ & 40  & $9.25^{+0.99}_{-0.89}$ & $0.686\pm0.064$ & 0.916 & 0.933 \\
$\lambda L_{\lambda}$(5100)/10$^{44}$  & 59 & $2.45^{+0.18}_{-0.17}$ & $0.608\pm0.045$ & 40  & $2.65^{+0.19}_{-0.18}$ & $0.694\pm0.056$ & 0.869 & 0.845 \\ 
                                       & 33 & $2.07^{+0.26}_{-0.23}$ & $0.665\pm0.065$ & 46  & $2.39^{+0.32}_{-0.28}$ & $0.723\pm0.075$ & 0.884 & 0.910 \\ 
$\lambda L_{\lambda}$(1450)/10$^{44}$  & 58 & $2.12^{+0.17}_{-0.15}$ & $0.496\pm0.042$ & 41  & $2.27^{+0.17}_{-0.16}$ & $0.595\pm0.046$ & 0.865 & 0.807 \\
                                       & 32 & $1.76^{+0.26}_{-0.23}$ & $0.545\pm0.063$ & 45  & $2.01^{+0.24}_{-0.21}$ & $0.604\pm0.053$ & 0.888 & 0.910 \\
$\lambda L_{\lambda}$(1350)/10$^{44}$  & 58 & $2.08^{+0.17}_{-0.16}$ & $0.499\pm0.040$ & 41  & $2.25^{+0.17}_{-0.16}$ & $0.595\pm0.046$ & 0.862 & 0.803 \\
                                       & 32 & $1.75^{+0.26}_{-0.22}$ & $0.540\pm0.056$ & 46  & $2.07^{+0.23}_{-0.21}$ & $0.584\pm0.046$ & 0.887 & 0.897 \\
$L$(2--10 keV)/10$^{43}$               & 54 & $0.75^{+0.11}_{-0.10}$ & $0.614\pm0.064$ & 52  & $0.65^{+0.13}_{-0.11}$ & $0.830\pm0.102$ & 0.705 & 0.669 \\
                                       & 30 & $0.86^{+0.18}_{-0.15}$ & $0.544\pm0.091$ & 64  & $0.71^{+0.24}_{-0.18}$ & $0.813\pm0.119$ & 0.688 & 0.701 \\
\cutinhead{Using only H$\beta$ time lag}
%
% \colhead{Luminosity}                 & N  &     A_ FITEXY          &   B_FITEXY      & EPS &     A_BCES             &     B_BCES      & Pears & Spear \\
$L$(H$\beta$)/10$^{43}$                & 55 & $7.55^{+0.91}_{-0.82}$ & $0.609\pm0.048$ & 38  & $9.42^{+1.02}_{-0.92}$ & $0.669\pm0.056$ & 0.874 & 0.845 \\
                                       & 30 & $7.18^{+0.94}_{-0.83}$ & $0.694\pm0.064$ & 40  & $8.79^{+1.00}_{-0.90}$ & $0.691\pm0.080$ & 0.906 & 0.923 \\
$L$(H$\beta$)/10$^{43}$ no narrow      & 55 & $7.85^{+0.94}_{-0.84}$ & $0.604\pm0.047$ & 38  & $9.73^{+1.10}_{-0.99}$ & $0.657\pm0.059$ & 0.867 & 0.854 \\
                                       & 30 & $7.49^{+0.99}_{-0.88}$ & $0.687\pm0.063$ & 40  & $9.18^{+1.09}_{-0.98}$ & $0.685\pm0.081$ & 0.906 & 0.925 \\
$\lambda L_{\lambda}$(5100)/10$^{44}$  & 55 & $2.43^{+0.18}_{-0.17}$ & $0.601\pm0.047$ & 39  & $2.69^{+0.20}_{-0.18}$ & $0.670\pm0.060$ & 0.870 & 0.843 \\
                                       & 30 & $2.00^{+0.28}_{-0.24}$ & $0.665\pm0.069$ & 44  & $2.47^{+0.36}_{-0.32}$ & $0.674\pm0.089$ & 0.882 & 0.911 \\
$\lambda L_{\lambda}$(1450)/10$^{44}$  & 54 & $2.11^{+0.17}_{-0.16}$ & $0.498\pm0.044$ & 39  & $2.28^{+0.17}_{-0.16}$ & $0.566\pm0.047$ & 0.858 & 0.780 \\
                                       & 29 & $1.72^{+0.29}_{-0.24}$ & $0.548\pm0.061$ & 44  & $2.13^{+0.29}_{-0.26}$ & $0.552\pm0.062$ & 0.878 & 0.894 \\
$\lambda L_{\lambda}$(1350)/10$^{44}$  & 54 & $2.06^{+0.18}_{-0.16}$ & $0.495\pm0.041$ & 40  & $2.26^{+0.17}_{-0.16}$ & $0.557\pm0.045$ & 0.854 & 0.776 \\
                                       & 29 & $1.72^{+0.27}_{-0.23}$ & $0.536\pm0.058$ & 44  & $2.14^{+0.27}_{-0.24}$ & $0.548\pm0.050$ & 0.876 & 0.878 \\
$L$(2--10 keV)/10$^{43}$               & 50 & $0.74^{+0.11}_{-0.10}$ & $0.612\pm0.063$ & 48  & $0.65^{+0.15}_{-0.12}$ & $0.806\pm0.108$ & 0.700 & 0.690 \\
                                       & 27 & $0.86^{+0.18}_{-0.15}$ & $0.532\pm0.090$ & 62  & $0.71^{+0.31}_{-0.22}$ & $0.773\pm0.139$ & 0.668 & 0.695 \\
\enddata
\tablecomments{For each luminosity in column 1 two rows are given:
first row is the fit results where multiple data sets were used for
each object, and the second row is the fit results where all data
sets per object were averaged. See text for details.}
\tablenotetext{a}{Results of the fit which correspond to equation 1.}
\tablenotetext{b}{Given in percent of the value.}
\end{deluxetable*}

\begin{figure*}
\centerline{\includegraphics[width=18.0cm]{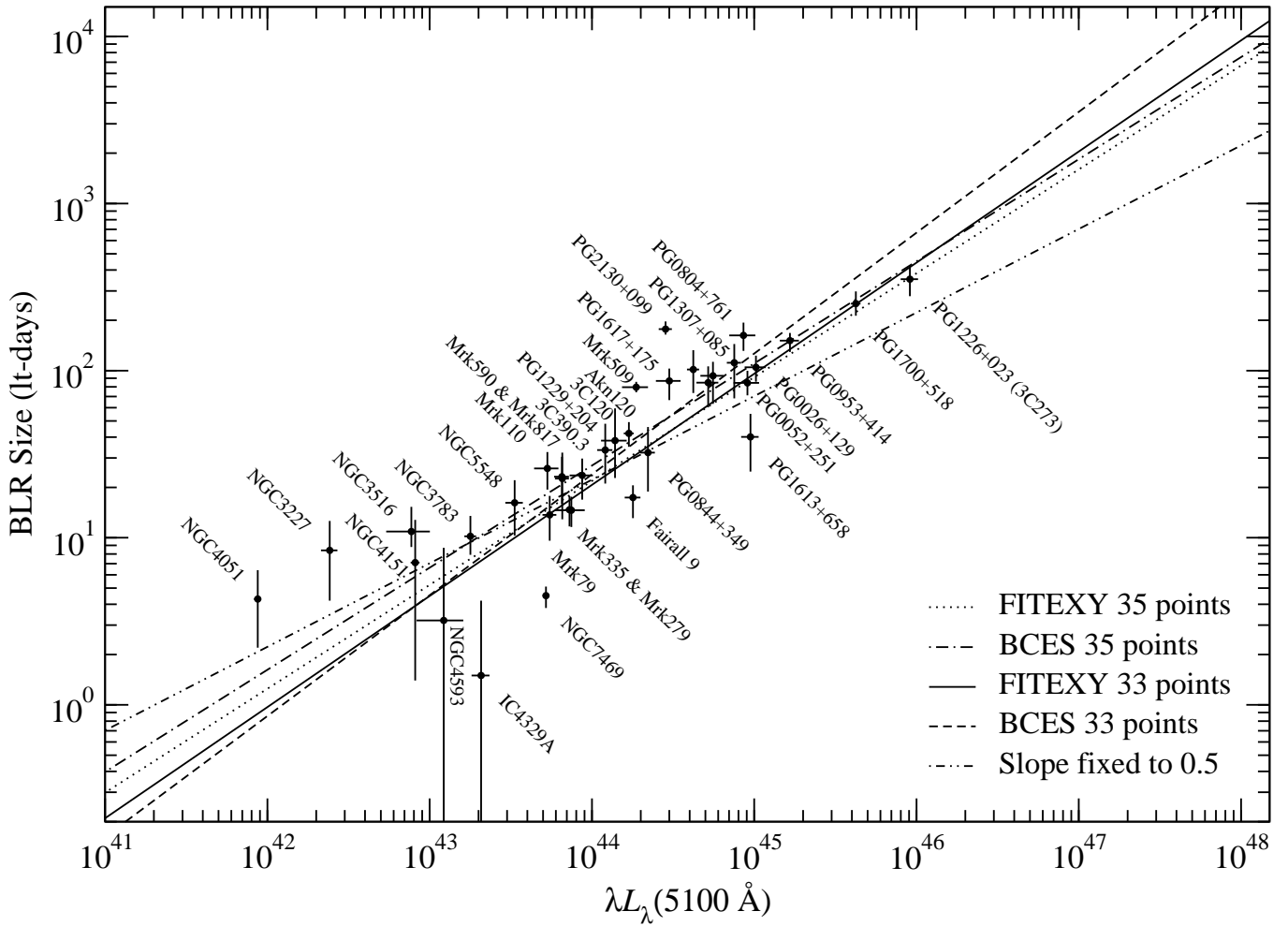}}
\caption{Balmer-line averaged BLR size plotted versus the $\lambda
L_{\lambda}$(5100\,\AA) luminosity (in units of \ergs ) on a larger
scale. The BLR size of each data set is determined from the averaged
Balmer-line time lags. Objects with multiple data sets have been
averaged to one point per object. Five line fits are shown: dotted
line - FITEXY using all 35 points; dot-dashed line - BCES using
all 35 points; solid line - FITEXY excluding the 2 lower luminosity
points; dashed line - BCES excluding the 2 lower luminosity points;
dot-dot-dashed line - FITEXY excluding the 2 lower luminosity
points and fixing the slope to 0.5.
\label{ext5100} }
\end{figure*}

In Figure~\ref{ext5100} we plot the mean Balmer-line BLR size versus
the $\lambda L_{\lambda}$(5100\,\AA ) luminosity, with one averaged
point per object. We also plot four fits: using all 35 points,
excluding the low luminosity AGNs, and with the two fitting methods.
Within the luminosity range of the measurements ($10^{43}$--$10^{46}$
\ergs) all fits are consistent with each other and all are well within
the scatter of the points in the plot. Also, when using only the data for
the more luminous ($\ga 5.2\times10^{43}$) objects (26 objects) from
Figure~\ref{ext5100}, we find a slope of B$_{\rm FITEXY}=0.63\pm0.07$
and a constant of A$_{\rm FITEXY}=2.4\pm0.3$, which are consistent
with the results when using the whole sample.

Comparing the two methods of analysis, using all data sets versus
using only one averaged data set per object, we find that the slope
parameter (B) is somewhat different while the normalization parameter
(A) is similar (both for the BCES and FITEXY methods). In general,
when using one averaged data point per object, the slopes are larger
by $\sim 0.05$ (which is within the formal uncertainty) than when
using all data sets. An argument for using one averaged data point
per object is that we do not know if the BLR-size -- luminosity
relation in a given object is the same as between different objects
(the physical processes driving the relation in a single AGN might
be different from the ones driving it in a range of objects).  Thus,
in order to have a relation that spans a large range of luminosity
and different objects and to study the physics across the luminosity
function, it is better to use the relation with a mean BLR-size and
a mean luminosity for each object. On the other hand, if the relation
in one object is the same as between objects there is justification to
use all data sets. This has been tested observationally for only one
object, NGC\,5548, by Peterson et al. (2002) who found the BLR-size --
luminosity relation has a slope of 0.95 when using optical luminosity
and 0.53 for the UV luminosity.\footnote{ The 0.53 slope was found by
scaling the relation for the optical luminosity (0.95) by the slope
of the relation between the UV and the optical luminosities, in which
Peterson et al. (2002) find to be 0.56. We note, however, that Gilbert
\& Peterson (2003) favored a slope of 0.67 for the relation between
the UV and the optical luminosities, in which case the BLR-size --
UV luminosity slope of Peterson et al. (2002) is then 0.64.} While
the slope with the UV luminosity is similar to the one we find in
Table~\ref{results}, the optical luminosity slope is not. Hence, it is
still not clear if the relation in a single object is the same as in
a sample of objects and we present both results in the current study.

Removal of the H$\beta$ narrow component from the total H$\beta$
luminosity affects the fit results only at the 1\% level with no
obvious general trend. Also, when correlating the BLR size with the UV
luminosity, the difference between using the 1350~\AA\ or the 1450~\AA\
bands is negligible (less than 2\% and well within the uncertainties),
as expected. The flux densities in these bands are very similar,
and in some cases identical.

Wu et al. (2004) correlated the BLR size with the H$\beta$ luminosity
for the objects of Kaspi et al. (2000) using the data given there.
The H$\beta$ luminosity can be considered a good indicator for the
AGN luminosity as the line is driven by the ionizing continuum.
It was noted by Kaspi et al. (2000) that the use of the H$\beta$
luminosity instead of the optical luminosity do not change their
correlation results (see their \S 4.2). Wu et al. (2004) use only
one data set per object and assume a different cosmology ($H_{0}=75$
km\,s$^{-1}$\,Mpc$^{-1}$ and q$_{0}=0.5$)\footnote{This different
cosmology would cause a slightly flatter slope for the BLR-size --
H$\beta$-luminosity relation compared with the cosmology used in the
current work by about 0.05, which is well within the uncertainty
range of the Wu et al. (2004) result.} and obtain a slope of
$0.68\pm0.11$. This result is consistent with the one we obtain in
our analysis (Table~\ref{results}).

\section{What Determines the BLR Size?}

A theoretical slope for the relation between the BLR size and the
luminosity, predicted by various early papers, is 0.5. This is based
on the assumption that, on average, all AGNs have the same ionization
parameter, BLR density, column density, and ionizing spectral energy
distribution (SED). A slope of 0.5 is also expected if the BLR size is determined by the dust sublimation radius (Netzer \& Laor 1993).
However, several studies have indicated that the SED is
luminosity dependent, and thus at least part of the assumptions above
are not justified (e.g., Mushotzky \& Wandel 1989; Zheng \& Malkan
1993; Puchnarewicz et al. 1996).  Here we use the {\it observed
relations} of the BLR size with the different luminosity bands to
study which band influences the BLR size the most.

\subsection{UV vs. Optical Continuum Luminosity}

Use of the H$\beta$ and $\lambda L_{\lambda}$(5100\,\AA )
luminosities in the BLR-size -- luminosity relation yields similar
slopes of about 0.69. The slope with the UV luminosity [both $\lambda
L_{\lambda}$(1450\AA ) and $\lambda L_{\lambda}$(1350\AA )] is lower by
about 0.1 than the slopes using the optical bands (this is manifested
both in the FITEXY method and the BCES method). The UV-based slope
is the closest to the theoretical expectation of 0.5 based on the
assumptions mentioned above.

The UV band (i.e., 1200--3000 \AA ) is close in wavelength to the
ionizing continuum which has a large effect on the responding BLR gas.
However, the UV luminosities are based on measurements that are
generally not simultaneous with the reverberation measurements, and
sometimes greatly separated in time.  The luminosity in the optical
band (i.e., 4000--8000 \AA ) might suffer from systematics which skew
the BLR-size -- luminosity relation. These systematics may be the
stellar contribution to the optical luminosity of the Seyfert galaxies
(which was not taken into account in this work), or a change of the
AGN SED with luminosity, which translates to different slopes in the
BLR-size -- luminosity relation. The change in SED is suggested by the
fact that brighter AGNs have flatter optical-UV continua, ranging from
an average slope ($\beta$ in $L_\nu \propto \nu^{\beta}$) of about
$-1$ for Seyferts, to about $-0.3$ for quasars (e.g., Neugebauer et
al. 1989; Schmidt \& Green 1983; Francis et al. 1991). This means that
quasars have a larger UV/optical flux ratio than Seyferts (a quasar
which is 10 times more luminous than a Seyfert in the optical band
will be more luminous than the same Seyfert by a larger factor in the
UV band ).  Thus, when using UV, rather than optical luminosities,
the ratio of luminosities between quasars and Seyferts grows and the
slope of the BLR size versus luminosity relation becomes shallower
as indicated by Table~\ref{results}.

Nonetheless, as already discussed and as we see below, potentially
better indicators of the ionizing continuum, such as the H$\beta$
luminosity, seem to contradict the finding that 0.5 is the ``correct''
slope.  For comparison we plot in Figure~\ref{ext5100}, showing the
relation between mean BLR size and optical luminosity, a FITEXY fit
to 33 data points with a slope fixed to 0.5. a line with a slope. The
best fit $A$ parameter (see equation~\ref{eqfit}) is 2.23.

We note in passing, that for individual AGNs the ratio of the UV
luminosity to the optical luminosity is observed to increase as the
object becomes brighter (e.g., Peterson et al. 2002; Gilbert \&
Peterson 2003). This can affect the BLR-size -- luminosity relation
for those AGN with multiple data sets. A possible prediction would be
that the flattest slopes might result from BLR-size -- luminosity
relations that were derived from the case when including all
individual data sets, rather than using the average for each AGN. 

\subsection{H$\beta$ luminosity}

Simple arguments based on recombination theory suggest that, in
low-density low-optical depth gas the H$\beta$ emission-line luminosity
is a good indicator of the ionizing continuum luminosity as the line
luminosity is driven by this continuum. The physical conditions in the
BLR are, however, known to be more complex (Netzer 1990 and references
therein) yet we expect a general linear dependence between the ionizing
flux and the H$\beta$ luminosity.  The slope of the BLR-size --
H$\beta$ luminosity relation found here is almost identical to the
slope found using $\lambda L_{\lambda}$(5100\,\AA ).  This is in accord
with studies that show the Balmer-line luminosity scales as the optical
luminosity between different AGNs (e.g., Yee 1980; Shuder 1981).
However the BLR-size -- H$\beta$ luminosity relation does not
agree with the slope obtained using the UV continuum. This seems to
suggest that the $\lambda 1450$ continuum is not the best indicator of
the ionizing continuum or, perhaps, that our understanding of the BLR
physics is incomplete, i.e., that the H$\beta$ line luminosity may not
depend linearly on the ionizing continuum due to optical depth and/or
collisional de-excitation effects in hydrogen lines. It has also been
argued that a major line driver, and a possible BLR-size regulator,
is the very high energy (X-ray) continuum. We do not favor this
explanation because the fractional energy in this continuum is small
compared with other wavelength bands. This is all supplemented by the
fact that the observed line response reflects changes in ionization
over a large volume and is geometry dependent in a way that
obscures the local, immediate line response to continuum variations.
For example, the covering factor and mean gas density of the BLR
gas might decrease with distance from the central ionizing source
(e.g., Kaspi \& Netzer 1999), and the non-linear response in the
inner BLR thus might contribute more than the response from the
outer parts.  Evidently, the BLR-size -- luminosity relationship is
rather complicated and depends on several factors.

\subsection{X-ray Continuum}

The BLR size versus X-ray luminosity relation shows a large scatter
(50--60\%), though a general trend can be found.  The two regression
methods give somewhat different results for the slope.  We are not able
to explain the reason for this difference though it is likely that it is
caused due to the large scatter and the way each method utilize the data.
The slope of the best fit relation is similar in the FITEXY method to
the slope from using the optical luminosity but in the BCES method
it is steeper by about 0.1--0.2 (the best fit slope is of order
$\sim 0.8$).  AGNs are known to
have large variability amplitudes in the X-ray band. For most
objects in our sample the X-ray fluxes were determined based on one
observation of the object which could well be away from its average
flux state. Furthermore, different objects in the sample were observed
with different instruments and telescopes having different effective
bandpasses, thus enlarging further the scatter between the objects. For
most objects, the X-ray luminosity measurement is not contemporaneous
with the time lag measurement (also during a monitoring campaign
of months and years the X-ray luminosity could change over a large
range) and this can add to the scatter in the relation. Nevertheless,
the larger slope for the X-ray luminosity may be real, since it is
in accordance with the recent result that $\alpha_{\rm ox}$ in AGNs
(a measure for the flux ratios between the X-ray and optical bands)
anticorrelates with the luminosity (e.g., Vignali et al. 2003). This
implies that when the optical luminosity increases the X-ray luminosity
increases by a smaller factor, and this can explain the larger slope
we see in the relation of the BLR size with the X-ray luminosity.

\section{Summary}

Peterson et al. (2004) recently compiled and analyzed in a uniform and
self-consistent way the BLR size from all available AGN reverberation
mapping data obtained over the past 15 years. In this paper, we
correlate the BLR size with the luminosity $L$, in the X-ray, UV,
and optical bands.  We investigate the correlation with two different
regression methods as well as using  multiple data sets for individual
objects versus only one averaged data set for each object. Though
small systematic differences exist between the different methods of
analysis, the results are generally consistent. Assuming a power-law
relation \rblr$\propto L^{\alpha}$, we find the mean best-fitting $\alpha$
is about $0.67\pm0.05$ for the optical continuum and the broad H$\beta$
luminosity, about $0.56\pm0.05$ for the UV continuum luminosity, and
about $0.70\pm0.14$ for the X-ray luminosity. These slopes
are a simple average of the slopes found from the different methods
using mean Balmer-lines time lags and the uncertainties represent the
standard deviations of the different results. We note that for the
X-ray luminosity the two methods give somewhat different slopes.
The detailed results for each method are given in Table~\ref{results}
and the preferred choice of slope and normalization should depend on
the application it is used for. Simply averaging the measurements
obtained from the BCES and FITEXY methods for the relation between
the BLR size and the optical luminosity at 5100 \AA , using one data
point per object (sixth line of Table~\ref{results}), we find:
\begin{equation}
\frac{R_{BLR}}{10 {\rm \ lt\,days}}= (2.23\pm0.21)\left(\frac{\lambda L_{\lambda}(5100\,{\mbox{\AA}} )}{\rm 10^{44} \ ergs\,s^{-1}} \right)^{0.69\pm0.05}  .
\end{equation}
When using all individual measurements for each object (fifth line 
of Table~\ref{results}) the mean slope for the two methods is
$0.65\pm0.04$ and the mean normalization is $2.55\pm0.13$. 
We also find that the fits indicate an intrinsic scatter of $\sim
40$\% in these relations. This is the first time this intrinsic
scatter has been quantified. 

Our results reflect on the naive theoretical predicted slope for
the relation between the BLR size and the luminosity of $\alpha = 0.5$
which is based on the assumption that all AGNs have the same ionization
parameter, BLR density, column density, and ionizing spectral energy
distribution (SED). The fact that for most energy bands the slope
is different from $\alpha = 0.5$ indicates that at least for some of
these characteristics the simple assumption is not valid and they
probably show an evolution along the luminosity scale.

\acknowledgments

We thank A. Laor, E. O. Ofek, and O. Shemmer for helpful discussions.
We are grateful for several valuable suggestions by the anonymous referee.
We gratefully acknowledge the financial support of the Israel
Science Foundation, grant 545/00, the Zeff Fellowship (S.~K.), and
the National Science Foundation through grant AST-0205964 to The Ohio
State University (B.~M.~P.).
This research has made use of the NASA/IPAC Extragalactic Database
(NED) which is operated by the Jet Propulsion Laboratory, California
Institute of Technology, under contract with the National Aeronautics
and Space Administration.

\end{document}